\documentstyle[prl, aps, multicol]{revtex}

\def\be{\begin{equation}}
\def\ee{\end{equation}}
\def\ba{\begin{eqnarray}}
\def\ea{\end{eqnarray}}

\begin{document}
\begin{multicols}{2}

\noindent
{\bf Comment on "Orbital Paramagnetism of\\ Electrons in Proximity to a
Superconductor" }

\vspace{0.5cm}

In a recent letter, Bruder and Imry (BI) \cite{bruderimry} address the
fascinating reentrance effect in proximity coupled normal-metal coated
superconducting cylinders discovered by Mota and co-workers\cite{mota}
and challenging theoretical understanding to date. At stake is the
explanation of a low-temperature paramagnetic reentrance of order
unity.  While the authors refrain from an explanation of the
experimental results, they claim to find a `significant
paramagnetic effect' which they strongly relate to the layout and
findings of the experiment.  Here, we show that the straightforward
calculation of the paramagnetic change in the susceptibility $\Delta
\chi/\chi$ along the lines of their analysis is by several orders of 
magnitude smaller than experimentally observed.


Consider a superconducting cylinder (radius $R$) coated with a
normal-metal layer of thickness $d\ll R$ in an axial field 
$B_z=\partial_x A$.  The task is to
determine the current-field {rela}\-tion in the normal layer 
$j(x) = \int dx' \, K(x-x') A(x')$ 
($x =$ radial coordinate, $j$ is the azimuthal current density) and
combine it with Maxwell's equation to find a {\it self-consistent}
solution of the screening problem.  The discussion of screening
requires the knowledge of the {\it dispersive} response function
$K(q)$: while in a London superconductor the response is local
\cite{Tinkham}, $K(q) = K_0 = -c/4\pi\lambda_{\rm \scriptscriptstyle L}^2$
($\lambda_{\rm \scriptscriptstyle L} = $ penetration depth), in a
Pippard superconductor \cite{Tinkham} with $K(q) \approx K_0 /(1+q\xi_0)$ 
and in proximity induced superconductivity of a clean normal metal 
\cite{Zaikin} with $K(q) = K_0 \delta(q)$ the response is non-local in general.


Bruder and Imry define the two quasi-particle populations of Andreev-
and glancing states. They assume that the Andreev states carry only 
diamagnetic currents, while the glancing states additionally give 
a paramagnetic correction. Rather than deriving the
$q$-dependent response of the two populations, BI restrict their
analysis to the $q=0$ mode and present the total currents $I_D$ and
$I_P$ as a function of the total flux $\Phi \approx 2\pi R A(q=0)$
through the cylinder. Invoking a perturbative 
analysis of screening under the assumption of local response with 
$K(q)=K(0)$ (notably {\it not} a straightforward assumption in the context
of proximity), BI arrive at the final result $I_P^s/I_M \sim 0.1
\sqrt{d/R}$, where the currents $I_P^s$ and $I_M$ denote the
para- and diamagnetic currents after screening ($\lambda_{\text eff}\equiv 
\lambda_L$ for local response). This result suggests a
`significant correction to the Meissner effect' which is small only by
the geometric parameter $\sqrt{d/R}$.


However, the ratio $I_P^s/I_M$ is not a measurable quantity.
Rather, we have to calculate the correction in the diamagnetic
susceptibility due to the paramagnetic response of the glancing
states.  This is easily accomplished within the local description
of BI {\it without} making any further assumptions than
those made in \cite{bruderimry}. We go over to the current {\it
densities} $j_D$ and $j_P$ and obtain the constitutive relation $j(x)
= j_D + j_P \approx -(c/4\pi\lambda_{\rm \scriptscriptstyle L}^2) 
(1 - 0.1 \sqrt{d/R}) A(x)$. The glancing states then produce the slightly
larger screening length $\lambda
\approx \lambda_{\rm \scriptscriptstyle L} (1+ 0.05\sqrt{d/R})$,
while the total current $I = L_z\int dx (j_D + j_P) = I_M$ remains
unchanged (there is no paramagnetic correction to the Meissner
current).  Accounting for the finite penetration depth, the
suceptibility $4 \pi \chi = -1+2\lambda_{\rm \scriptscriptstyle L}/R$
changes by $\Delta \chi /\chi \approx - 0.1 (\lambda_{\rm \scriptscriptstyle
L}/R)\sqrt{d/R}$ when the paramagnetic response of the glancing
states is accounted for. This result is parametrically small by the
microscopic ratio $\lambda_{\rm \scriptscriptstyle L}/R$ and at best
amounts to a relative change $\Delta \chi /\chi \sim 10^{-4}$ when
using parameters appropriate to the experiments 
($\lambda_{\rm \scriptscriptstyle L}= (mc^2/4\pi n e^2)^{1/2}= 200 {\rm \AA}$, 
$R=20 \mu{\rm m}$, $d=5 \mu {\rm m}$) \cite{mota},
a result far from the experimental finding $\Delta \chi /\chi \sim 1$. 


We would like to point out that 
the separation of the quasi-particles into Andreev- and glancing
states has been used before by Belzig \cite{Belzig},
who, using quasi-classical techniques and accounting for the finite-$q$ 
paramagnetic response of the Andreev states, found a significant 
paramagnetic effect (of order $10^{-1}$, depending on geometry) 
in very clean material. The work of Belzig fails to
explain the experiments as it cannot produce the non-monotonic
temperature dependence crucial to the reentrance effect.
In this context, BI's reference to the temperature dependence 
of persistent currents does not account for two facts: This temperature
dependence is specific to the {\it quantum coherence} of the persistent
current states and cannot be trivially generalized to the {\it classical}
glancing states. Second, not only is the number of persistent current 
states parametrically small as compared to the number of glancing states,
the response of the persistent current states also is sample dependent 
and may be dia- or paramagnetic.


In conclusion, the work of Bruder and Imry does not address the 
salient points of the paramagnetic reentrance effect, 
which are its puzzling magnitude and non-monotonic 
temperature dependence --- the experiment still waits for an
explanation.\\

\noindent
{\it A.\ Fauch\`ere, V.~Geshkenbein, and G. Blatter\\
Theoretische Physik, ETH H\"onggerberg,\\ CH-8093 Z\"urich}

\vfill

\end{multicols}
\end{document}